\documentclass[twocolumn, deluxetables]{aastex631}
\usepackage{amsmath}
\usepackage{multirow}
\usepackage{graphicx}

\graphicspath{{./}{figures/}}

\shorttitle{XRISM Spectroscopy of 4U 1630-472}
\shortauthors{Miller, Mizumoto, Shidatsu, et al.}

\begin{document}

\title{XRISM Spectroscopy of the Stellar-Mass Black Hole 4U 1630$-$472 in Outburst}

\author[0000-0003-2869-7682]{Jon M. Miller}
\email{jonmm@umich.edu}
\affiliation{Department of Astronomy, University of Michigan, Ann Arbor, MI, 48109, USA}

\author[0000-0003-2161-0361]{Misaki Mizumoto}
\affiliation{Science Research Education Unit, University of Teacher Education Fukuoka, Akama-bunkyo-machi, Munakata, Fukuoka 811-4192, Japan}

\author[0000-0001-8195-6546]{Megumi Shidatsu}
\affiliation{Department of Physics, Ehime University, Ehime 790-8577, Japan} 

\author[0000-0002-1118-8470]{Ralf Ballhausen}
\affiliation{Department of Astronomy, University of Maryland, College Park, MD 20742, USA} 
\affiliation{X-ray Astrophysics Laboratory, NASA / Goddard Space Flight Center (GSFC), Greenbelt, MD 20771, USA}
\affiliation{Center for Research and Exploration in Space Science and Technology, NASA / GSFC (CRESST II), Greenbelt, MD 20771, USA}

\author[0000-0001-9735-4873]{Ehud Behar}
\affiliation{Department of Physics, Technion, Technion City, Haifa 3200003, Israel}

\author[0000-0001-7796-4279]{Mar{\'i}a D{\'i}az Trigo}
\affiliation{ESO, Karl-Schwarzschild-Strasse 2, 85748, Garching bei München, Germany}

\author[0000-0002-1065-7239]{Chris Done}
\affiliation{Centre for Extragalactic Astronomy, Department of Physics, University of Durham, South Road, Durham DH1 3LE, UK}

\author{Tadayasu Dotani}
\affiliation{Institute of Space and Astronautical Science (ISAS), Japan Aerospace Exploration Agency (JAXA), Kanagawa 252-5210, Japan} 

\author[0000-0003-3828-2448]{Javier A. Garc{\' i}a}
\affiliation{Cahill Center for Astronomy and Astrophysics, California Institute of Technology, 1200 E. California Boulevard, MC 290-17, Pasadena, CA 91125, USA} 

\author[0000-0002-5779-6906]{Timothy Kallman}
\affiliation{X-ray Astrophysics Laboratory, NASA / Goddard Space Flight Center (GSFC), Greenbelt, MD 20771, USA}

\author[0000-0001-7773-9266]{Shogo B. Kobayashi}
\affiliation{Faculty of Physics, Tokyo University of Science, Tokyo 162-8601, Japan}

\author{Aya Kubota}
\affiliation{Department of Electronic Information Systems, Shibaura Institute of Technology,  Saitama 337-8570, Japan}

\author[0000-0003-4284-4167]{Randall Smith}
\affiliation{Center for Astrophysics | Harvard-Smithsonian, MA 02138, USA} 

\author[0000-0001-6314-5897]{Hiromitsu Takahashi}
\affiliation{Department of Physics, Hiroshima University, Hiroshima 739-8526, Japan} 

\author[0000-0002-5097-1257]{Makoto Tashiro}
\affiliation{Department of Physics, Saitama University, Saitama 338-8570, Japan} 
\affiliation{Institute of Space and Astronautical Science (ISAS), Japan Aerospace Exploration Agency (JAXA), Kanagawa 252-5210, Japan} 

\author[0000-0001-7821-6715]{Yoshihiro Ueda}
\affiliation{Department of Astronomy, Kyoto University, Kyoto 606-8502, Japan} 

\author[0000-0002-4708-4219]{Jacco Vink}
\affiliation{Anton Pannekoek Institute/GRAPPA, University of Amsterdam, Science Park 904, 1098 XH Amsterdam, The Netherlands}
\affiliation{SRON Netherlands Institute for Space Research, Niels Bohrweg 4, 2333 CA Leiden, The Netherlands}

\author[0000-0003-4808-893X]{Shinya Yamada}
\affiliation{Department of Physics, Rikkyo University, Tokyo 171-8501, Japan} 

\author[0000-0003-0441-7404]{Shin Watanabe}
\affiliation{Institute of Space and Astronautical Science (ISAS), Japan Aerospace Exploration Agency (JAXA), Kanagawa 252-5210, Japan} 

\author{Ryo Iizuka}
\affiliation{Institute of Space and Astronautical Science (ISAS), Japan Aerospace Exploration Agency (JAXA), Kanagawa 252-5210, Japan} 

\author[0000-0002-2359-1857]{Yukikatsu Terada}
\affiliation{Department of Physics, Saitama University, Saitama 338-8570, Japan} 
\affiliation{Institute of Space and Astronautical Science (ISAS), Japan Aerospace Exploration Agency (JAXA), Kanagawa 252-5210, Japan} 

\author{Chris Baluta}
\affiliation{National Aeronautics and Space Administration (NASA), Goddard Space Flight Center, 8800 Greenbelt Road, Greenbelt, Maryland, 20771, United States}

\author[0000-0002-4541-1044]{Yoshiaki Kanemaru}
\affiliation{Institute of Space and Astronautical Science (ISAS), Japan Aerospace Exploration Agency (JAXA), Kanagawa 252-5210, Japan} 

\author[0000-0002-5701-0811]{Shoji Ogawa}
\affiliation{Institute of Space and Astronautical Science (ISAS), Japan Aerospace Exploration Agency (JAXA), Kanagawa 252-5210, Japan} 

\author{Tessei Yoshida}
\affiliation{Institute of Space and Astronautical Science (ISAS), Japan Aerospace Exploration Agency (JAXA), Kanagawa 252-5210, Japan} 

\author[0000-0001-6922-6583]{Katsuhiro Hayashi}
\affiliation{Institute of Space and Astronautical Science (ISAS), Japan Aerospace Exploration Agency (JAXA), Kanagawa 252-5210, Japan} 

\begin{abstract}
We report on XRISM/Resolve spectroscopy of the recurrent transient and well-known black hole candidate 4U~1630$-$472 during its 2024 outburst.  The source was captured at the end of a disk-dominated high/soft state, at an Eddington fraction of $\lambda_\mathrm{Edd} \sim 0.05~(10~M_{\odot}/M_\mathrm{BH})$.  A variable absorption spectrum with unprecedented complexity is revealed with the Resolve calorimeter.  This marks one of the lowest Eddington fractions at which highly ionized absorption has been detected in an X-ray binary.  The strongest lines are fully resolved, with He-like Fe XXV separated into resonance and intercombination components, and H-like Fe XXVI seen as a spin-orbit doublet.    The depth of some absorption lines varied by almost an order of magnitude, far more than expected based on a $10$\% variation in apparent X-ray flux and ionization parameter.  The velocity of some absorption components also changed significantly.  Jointly modeling two flux segments with a consistent model including four photoionization zones, the spectrum can be described in terms of highly ionized but likely failed winds that sometimes show red-shifts, variable obscuration that may signal asymmetric structures in the middle and outer accretion disk, and a tentative very fast outflow ($v = 0.026-0.033c$).  We discuss the impact of these findings on our understanding of accretion and winds in stellar-mass black holes, and potential consequences for future studies.
\end{abstract}


\section{Introduction}
Across the mass scale, accretion onto black holes is accompanied by mass loss in the form of winds (for a review on black hole feedback, see \citealt{fabian2012}; also see \citealt{gmc2023}).  In active galactic nuclei (AGN), relatively low gas ionization parameters may make it possible for radiation pressure on UV lines to drive winds (see, e.g., \citealt{proga2000}), including the extreme flows seen in broad absorption line quasars (BALQSOs) and potentially ultra-fast outflows in X-rays (UFOs: $v/c \geq 0.1$; see \citealt{tombesi2010}).  These winds may carry the kinetic power necessary to reshape host galaxy bulges, roughly 0.5--5\% of the bolometric luminosity (\citealt{tdm2005}, \citealt{hopkins2006}).  In contrast, the X-ray disk winds observed in stellar-mass black holes are generally more highly ionized, and different mechanisms likely drive the extreme winds in these environments (for a review, see \citealt{neilsen2023}).  Understanding X-ray disk winds in stellar-mass black holes is important to developing an accurate picture of binary evolution, and even the fundamental physics of disk accretion.

Observations of stellar-mass black holes with current X-ray missions have unraveled some key facets of disk winds in stellar-mass black holes.  Disk winds are preferentially observed in spectrally soft, disk--dominated states, and in systems wherein the line of sight passes close to the plane of the accretion disk (\citealt{miller2008}, \citealt{neilsen2009}, \citealt{king2012}, \citealt{ponti2012}; also see \citealt{shidatsu2013}, \citealt{parra2024}). These findings indicate that disk winds are equatorial, at least in the sense that most of the
gas column density is confined to regions just above the plane of the
underlying optically thick disk.  Changes in the gas ionization, column
density, or magnetic field configuration have been discussed as the
reasons for inhibiting wind production or making it invisible via
over-ionization in the spectrally hard states. Distinguishing among those
possibilities could also provide further clues to the origin of winds in
X-ray binaries and the overall mass outflow budget during the outburst.

For this reason, many recent efforts have been dedicated
to determine the precise moment at which winds are launched and quenched
and on the conditions (luminosity, spectral state or jet presence)
prevalent at such epochs. A transition from a soft state to a very high
state of 4U 1630$-$47 was extensively monitored in its 2012 outburst.
The wind ionization increased until it disappeared,
just before the state transition \citep{diaztrigo2014}. The source was also observed during the outburst decay with Chandra, but no lines were detected in any of the spectra, leading to the possibility that the wind may
have been depleted before the transition occurs \citep{gatuzz2019}.

The most complex stellar-mass black hole disk winds have been observed in GRO~J1655$-$40 and GRS~1915$+$105 using the Chandra High Energy Transmission Grating Spectrometer (HETGS).  The best models for the deepest HETGS data find four distinct wind zones in each source, with velocities spanning from a $\mathrm{few} \times 10^{2}~{\rm km}~{\rm s}^{-1}$ to $\mathrm{few} \times 10^{4}~{\rm km}~{\rm s}^{-1}$, and ionization parameters that also span roughly two orders of magnitude (\citealt{miller2015}, \citealt{miller2016}; also see \citealt{balakrishnan2020}). The broad range of values may suggest that the winds themselves are driven over a fairly broad range in radius, and potentially a combination of physical mechanisms.  A plausible scenario is that magnetic pressure may act at high velocity and high ionization and/or at small radii, whereas thermal pressure can act at large radius and yield low velocity flows (\citealt{neilsen2012}; also see \citealt{keshet2024}).  The magnetic field required to drive the disk wind observed in the soft state of GRS~1915$+$105 is compatible with expectations for a standard, Shakura-Sunyaev disk \citep{ss73}, potentially suggesting that such winds are a natural product of thin disk accretion \citep{miller2016}.

Absorption lines with no significant shifts are sometimes observed in X-ray binaries with similar line-of-sight angles to those with outflowing disk winds \citep{shidatsu2013}, suggesting a static structure with a relatively small scale height above the disk. This may be compatible with a hot atmosphere above the disk \citep{rozanska2014}.  X-ray irradiation forms photoionized plasma on the disk surface, which is kept bound as a hot atmosphere inside the minimum radius to launch a thermal wind \citep{jimenezgarate2002}, if no magnetic wind is present.  Such a hot atmosphere was found to well explain the static absorption lines seen in many low-mass X-ray binaries hosting a neutron star, which often have a relatively small accretion disk compared to those 
in black hole X-ray binaries (\citealt{diaztrigo2016}; but note the fast disk wind in 4U 1916$-$053, \citealt{trueba2020}). Another possible ionized absorber is the outer disk structure that is responsible for absorption dips (e.g., \citealt{boirin2005}; \citealt{diaztrigo2006}; \citealt{shidatsu2013}), which is formed on the outer disk near the impact point of the accretion stream from the companion star. It usually produces denser and lower ionized absorption than those of a disk wind and a hot atmosphere. Static absorption lines provide information of these structures, such as the location, geometry, density, and ionization properties, and thus they are a good probe of the accretion disk and its surrounding structure outside the X-ray emitting region. 

Whereas GRO~J1655$-$40 has only had one outburst since the launch of Chandra, and whereas GRS~1915$+$105 has been in a faint, heavily obscured state for years (see, e.g., \citealt{miller2020}, \citealt{balakrishnan2021}), 4U 1630$-$472 is a candidate stellar-mass black hole that enters a bright outburst phase every two years \citep{capitanio2015}.  It typically reaches a high/soft state and launches a disk wind, making it the best candidate for improving our understanding of disk winds and connected disk physics in stellar-mass black holes (see, e.g., \citealt{gatuzz2019}).  The mass of the primary in 4U~1630$-$472 has not been determined using optical spectroscopy because the line of sight column density is roughly $N_{H} = 10^{23}~{\rm cm}^{-2}$ (e.g., \citealt{miller2015}), translating to $A_{V} = 45$ \citep{guver2009}.  However, it is well-known for high-frequency quasi-periodic oscillations typical of stellar-mass black holes \citep{kw04}, relativistic disk reflection and a spin value that nominally excludes a neutron star primary (\citealt{king2014}, \citealt{draghis2024}), and other phenomena that strongly indicate that the binary harbors a black hole. 
\citet{kalemci2018} estimated the distance of the source to be $11.5 \pm 0.3$ kpc, from the analysis of the dust-scattering halo produced during its outburst, although another possibility -- $4.7 \pm 0.3$ kpc -- was not completely ruled out.  

Suzaku observations of 4U~1630$-$472 revealed blue-shifted Fe~XXV and Fe~XXVI absorption lines with a speed of $v \sim -1000~{\rm km}~{\rm s}^{-1}$, likely produced at a radius of just $r \sim 10^{10}$~cm \citep{kubota2007}.  Interestingly, the more highly ionized Fe XXVI line was found to be roughly constant in equivalent width across six observations, while the Fe XXV line equivalent width roughly tripled as the ionizing flux decayed.  This may have been an early indication of wind multiple zones and driving mechanisms in 4U 1630$-$472.  Later work interpreted new but similar Suzaku observations of 4U~1630$-$472 only in terms of a thermal wind \citep{hori2018}.  In contrast, NuSTAR spectroscopy of the wind in 4U 1630$-$472 find evidence of a fast, $v = 0.04c$ wind launched from just $r = 200-1100~GM/c^{2}$, requiring magnetic driving \citep{king2014}.  

The resolution afforded by Chandra may provide the best prior view of disk winds in 4U~1630$-$472.  \citet{miller2015} fit a particularly sensitive HEG spectrum (ObsID 13715), using high-resolution XSTAR models.  Two fairly different wind components were revealed: a slower, lower-ionization component with a very high column density ($v = 200~{\rm km}~{\rm s}^{-1}$, log$\xi = 4.1$, $N_H = 2.2\times 10^{23}~{\rm cm}^{-2}$, and $v = 1800~{\rm km}~{\rm s}^{-1}$, log$\xi = 4.6$, and $N_H = 7\times 10^{22}~{\rm cm}^{-2}$; also see \citealt{gatuzz2019}).  Nominally, these zones are consistent with a slow thermal wind from the outer disk, and a magnetic wind from the inner disk.  However, this modeling assumed that both components see the same ionizing flux.  \citet{trueba2019} modeled numerous Chandra HEG spectra of 4U~1630$-$472 using the ``pion'' model in SPEX, which allows absorbers to be layered.  Absorbers that are exterior to inner layers of absorption see a modified continuum, with a lower total flux.  When this effect is taken into account, the slower wind components are found to also originate at relatively small radii, and the distribution of wind column density with ionization -- the absorption measure distribution (or, AMD) -- is found to be relatively flat and consistent with magnetic driving (similar results are reported in GRO~J1655$-$40; \citealt{balakrishnan2020}, \citealt{keshet2024}).

In this paper, we analyze the disk wind spectrum revealed in a XRISM observation of 4U~1630$-$472, obtained at a lower flux than most prior studies.  The results directly impact our understanding of disk wind evolution on long and short time scales, vertical and azimuthal structure, and launching mechanisms.  In Section 2, we present the observation details and data reduction.  Our analysis and results are presented in Section 3.  These findings are discussed in Section 4.  Finally, we restate our findings as a concise set of conclusions in Section 5. Throughout the paper, we assume a distance of 11.5 kpc and a black hole mass of $10 M_\odot$.

\section{Observations and Data Reduction}
\begin{figure*}
 \begin{center}
  \includegraphics[width=1.0\columnwidth]{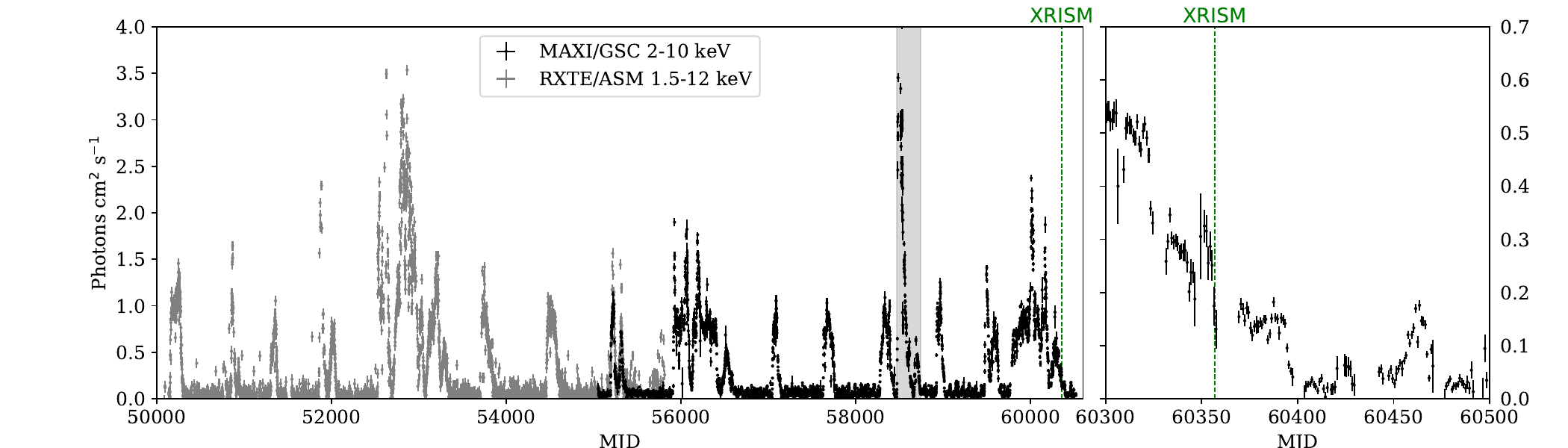} 
 \end{center}
\caption{(Left) long-term X-ray light curve of 4U 1630$-$472 over $\sim 30$ years, obtained with the RXTE/ASM (gray) and the MAXI/GSC (2--20 keV; black). The brightening in the shaded region is due to the nearby source MAXI J1630$-$276. 
The RXTE/ASM intensity is normalized so that the light curve profile of the outburst around MJD 55200--55400 fits that of the MAXI/GSC. 
(Right) enlarged view around the XRISM observation.}
\label{fig:lc_longterm}
\end{figure*}

Figure~\ref{fig:lc_longterm} shows the long-term light curve of 4U 1630$-$472 obtained by combining the MAXI/GSC and the RXTE/ASM data\footnote{downloaded from \url{http://maxi.riken.jp/star_data/J1634-473/J1634-473.html} for MAXI and \url{https://heasarc.gsfc.nasa.gov/FTP/xte/data/archive/ASMProducts/definitive_1dwell/lightcurves/} for RXTE.}. The source repeatedly shows several relatively small outbursts, and some larger outbursts. 
The XRISM observation (OBSID: 900001010) was conducted in the last giant outburst as of 2024, at the last part of its decaying phase, as the first ToO observation of the XRISM mission. The observation started on 16 February 2024 at 01:31:50.96 UTC and ended on 17 February at 05:01:58 UTC. During the observation, Resolve was operated in the open-filter mode.  Xtend data are important for spatially extended sources, and in cases where absorption in the Milky Way does not scatter away the low energy source flux.  This initial paper on 4U~1630$-$472 is therefore restricted to analysis of the Resolve data.

The Resolve observation was performed through a closed gate valve, which is opaque in $E\lesssim1.8$~keV \citep{Midooka2021}.
Good Time Interval (GTI) filtering is applied to exclude periods of the Earth eclipse and sunlit Earth's limb, South Atlantic Anomaly (SAA) passages, and times within 4300~s of the beginning of a recycling of the 50-mK cooler. 
Data reduction was conducted using pre-pipeline software version JAXA ``004\_002.15Oct2023\_Build7.011'', along with the pipeline script version ``03.00.011.008,'' and the internal CALDB8, aligned with the public XRISM CALDB version 20240815. Event screening utilized anti-coincidence (anti-co) detector information, as well as coincidence data from other Resolve pixels. Pixel-to-pixel coincidence screening was primarily implemented to exclude multi-pixel events induced by cosmic rays depositing energy into the frame around the pixels \citep{Kilbourne2018}; an energy threshold of 300 eV was set to mitigate the inclusion of crosstalk events. Additionally, a standard rise time cut, adjusted for energy dependence, was applied as per established methodology \citep{mochizuki2024}.
After filtering, the net Resolve exposure time is 50.29~ks.  A total count rate of $14.0~{\rm counts}~{\rm s}^{-1}$ is recorded, or $13.8~{\rm counts}~{\rm s}^{-1}$ in the canonical 2--10~keV band.  The rates are similar because of the high ISM column density along the line of sight to 4U~1630$-$472 and a thermal spectrum that falls rapidly above 10~keV.

Calorimeter events carry ``grades'' based on their energy resolution.  At high count rates, or in periods of high background, a fraction of events may have slightly lower resolution.  The count rates observed from 4U~1630$-$472 are modest and the vast majority of events are high-resolution primary or ``Hp'' events.  In order to take advantage of the best possible spectral resolution, then, this analysis is confined to spectra made from ``Hp'' events.

\begin{figure}
    \centering
    \includegraphics[width=0.5\columnwidth]{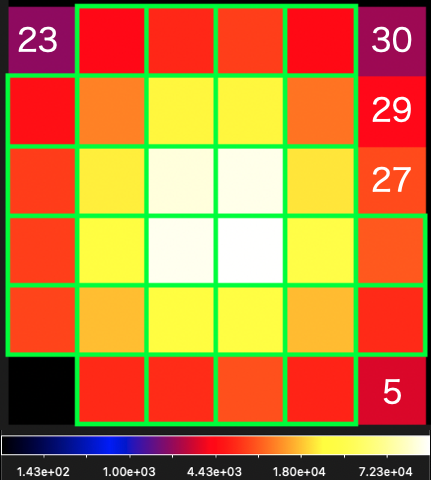}
    \caption{The Resolve image in the detector coordinates. We do not use Pixel 5, 23, 29, 30 due to pseudo-Ls event pollution, or Pixel 27 due to the possible temperature jump.}
    \label{fig:image}
\end{figure}

For Resolve data, many events that saturate the output of the signal digitizer (produced by shallow-angle cosmic rays) get flagged as one Lp (low-resolution primary) and multiple false Ls (low-resolution secondary) events through the action of the secondary-pulse detection algorithm on a clipped pulse. In the pixels with low count rate, the fraction of false-Ls events can be relatively large, the Ls fraction can be overestimated, and the estimated flux from the Hp spectrum can be underestimated. We have found that 4 pixels (Pixel 5, 23, 29, 30; see Figure \ref{fig:image}) at the outer edge of the array have count rates of $<0.08$~s$^{-1}$~pix$^{-1}$ and have a relatively higher Ls fraction than estimated. Therefore, we have removed the 4 pixels from the subsequent analysis.

Time-dependent gain tracking for each pixel was performed using the Mn K$\alpha$ lines from a set of \(^{55}\)Fe sources on the filter wheel. The lines intermittently illuminated the whole array. Our observations include 5 gain fiducial measurements: just after the start of the observation, before and after the 50~mK cooler recycle, about 3 hours after the recycle, and just before the end of the observation. It is known that Pixel 27 randomly shows temperature jumps with a duration of $\sim$1--2 hours. Its origin is unknown, and the current gain fiducial measurement strategy cannot track these temperature jumps. Therefore we determined to exclude Pixel 27. Consequently, we use 30 pixels for our analysis, as shown in Figure \ref{fig:image}.
Gaussian modeling of the detector line spread resulted in an instrument resolution of $4.44\pm0.07$~eV in Full-Width Half Maximum (FWHM), with an energy shift evaluated at less than 0.1 eV. Early calibration measurements indicate that the energy scale is accurate to within approximately $\pm1$~eV from 2--8 keV \citep{porter2024,eckart2024}.

A redistribution matrix file (RMF) was generated via the \texttt{rslmkrmf} task using the cleaned event file without any Ls events and CALDB based on ground measurements. The following line-spread function components were considered: the Gaussian core, exponential tail to low energy, escape peaks, silicon fluorescence, and electron loss continuum. In other words, the ``X'' option was adopted. An auxiliary response file (ARF) was generated using the \texttt{xaarfgen} task, assuming a point-like source at the aim point as input.

To accurately calculate the flux from photon counts, it is essential to determine the correct branching ratio of the grade, which varies with the incoming rate \citep{Ishisaki2018}. As explained in Section 3.1, the target showed a flux variation during observation and, we divided the observation into  two phases based on this flux change. Separate response files were generated for each phase, and model fitting was performed accordingly. The background is negligible.

\section{Analysis and Results} 
\subsection{Light Curves}

\begin{figure*}
 \begin{center}
  \includegraphics[width=1.0\columnwidth]{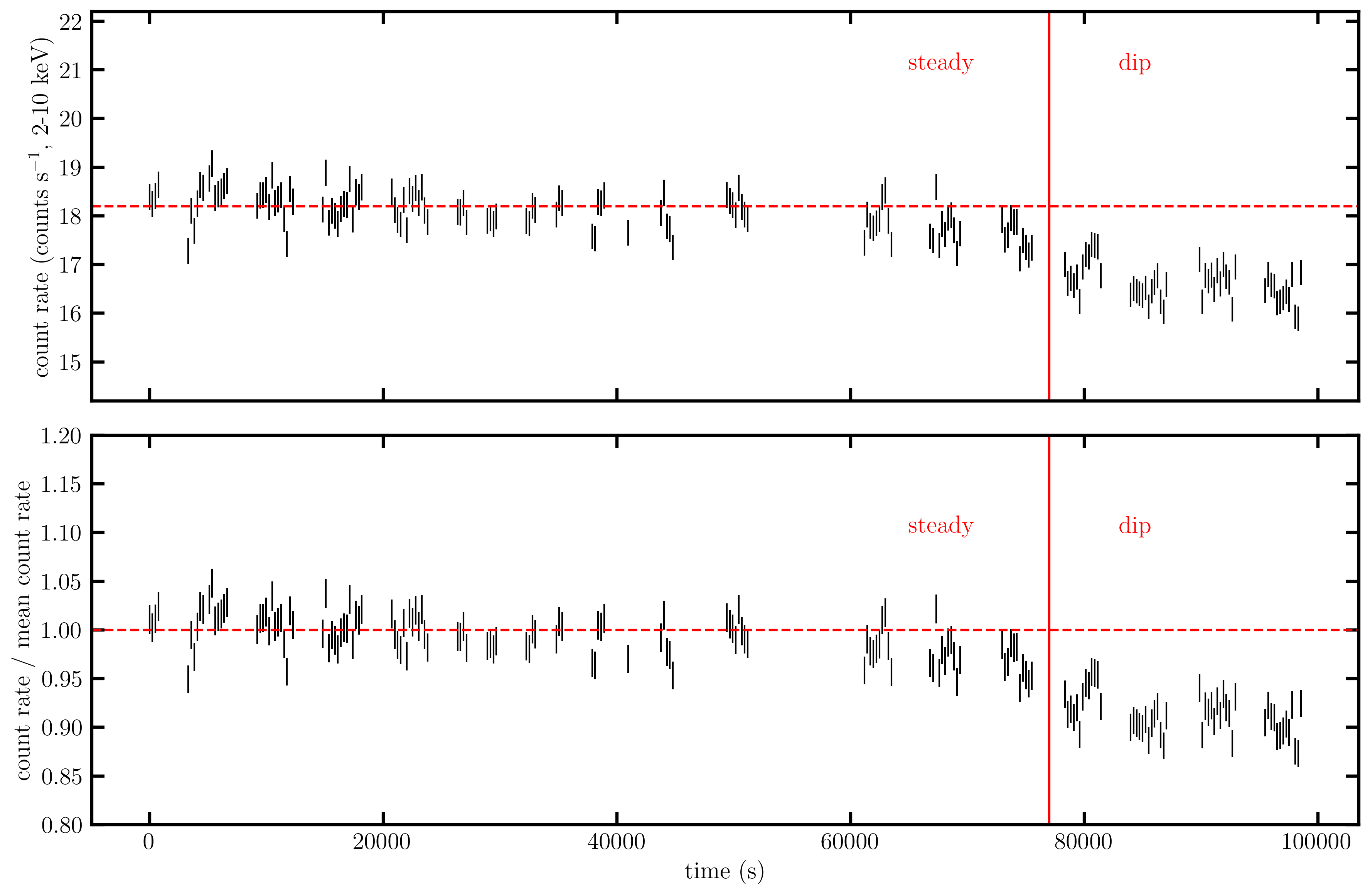} 
 \end{center}
\caption{The Resolve light curve of 4U~1630$-$472 in the 2--10~keV band, with 256~s bins.  The dashed horizontal red lines in each panel show the mean count rate prior to $t=60,000$~seconds.  The solid vertical line in each panel marks the point at which the spectra are divided into ``steady'' flux and ``dip'' phases.}
\label{fig:lc_resolve}
\end{figure*}

Figure~\ref{fig:lc_resolve} shows the Resolve light curve in the 2--10 keV band. The source intensity was almost constant for the first $\sim 7 \times 10^4$ s, and then gradually decreased.  The flux became fairly stable again at $\sim 90$\% of the initial intensity level.  This behavior is potentially consistent with a broad ``dip,'' so we divided the observation period around the flux decrease as indicated in Fig.~\ref{fig:lc_resolve}.  We made separate spectra from these time segments, referred to as the ``steady'' and ``dip'' phases in the following sections.

\subsection{Spectra}
\subsubsection{Continuum}
\begin{figure*}
 \begin{center}
  \includegraphics[width=1.0\columnwidth]{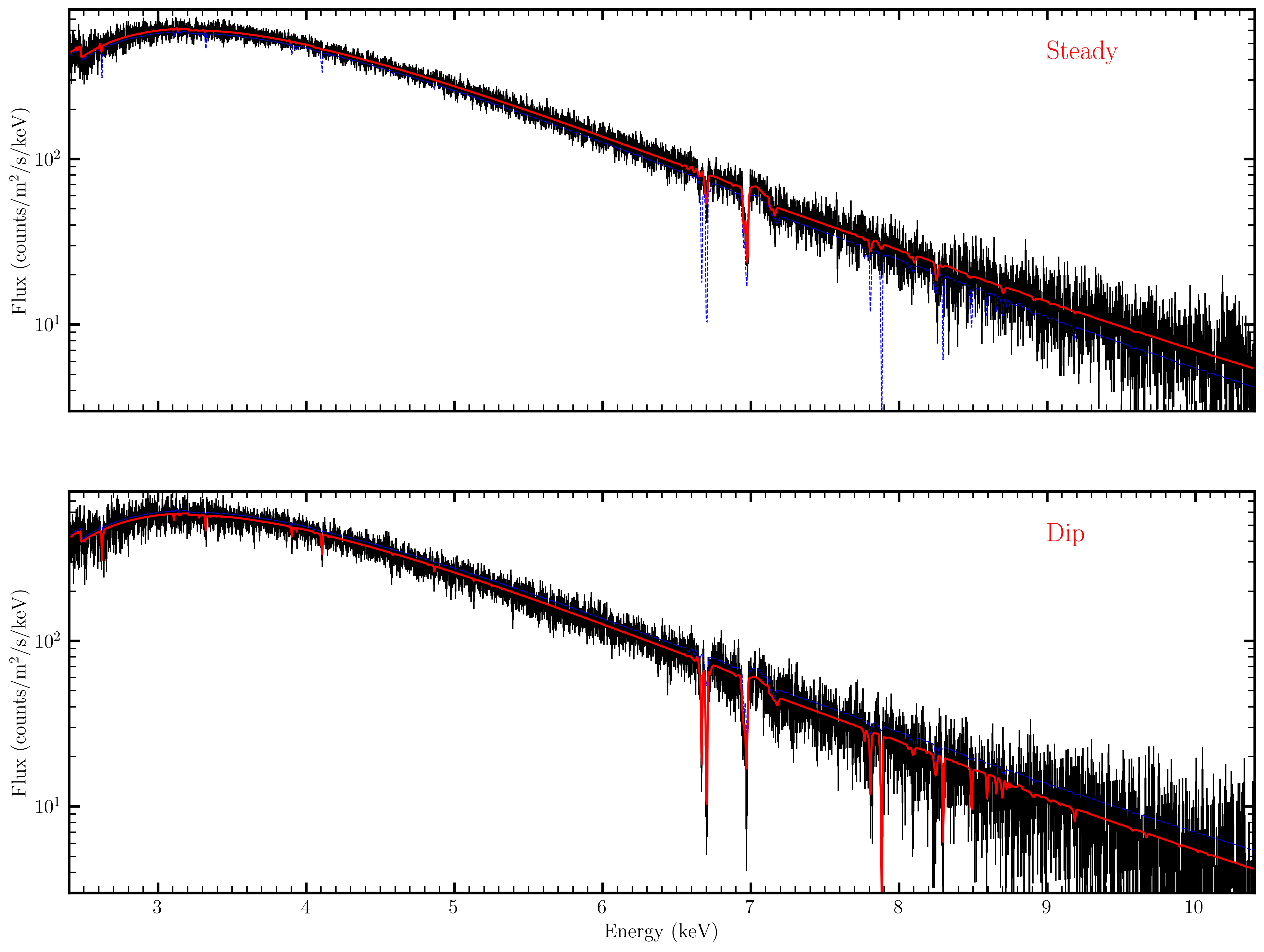} 
 \end{center}
\caption{Resolve spectra of 4U~1630$-$472 obtained from the initial ``steady'' flux phase of the observation, and the latter ``dip'' phase.  The spectra are dominated by narrow He-like Fe XXV and H-like Fe XXVI absorption lines.  In both phases, the best-fit ``pion'' photoionization model is shown in red; it includes four independent absorption and re-emission zones.  In each panel, the best-fit to the complementary phase is indicated with a dashed blue line.
The phases are primarily distinguished by a sharp increase in absorption that cannot be accounted for only through a change in luminosity and corresponding change in ionization.  The spectra were binned using the ``optimal binning'' algorithm and are shown at the same resolution.  Please see the text and subsequent figures for additional details.}
\label{fig:best_broad}
\end{figure*}

In order to derive the simplest overall model that could represent similarities and differences, spectra from the ``steady'' and ``dip'' phases were fit jointly within SPEX.  Prior to fitting, the spectra were binned using the ``optimal' binning algorithm \citep{kaastra2016}, minimizing a Cash statistic \citep{cash1979}.  The high line-of-sight column density to 4U~1630$-$472 results in poor sensitivity close to the gate valve boundary, and calibration of that region is ongoing.  We therefore set a lower energy threshold of 2.4~keV for our fits.  The overall spectrum is very soft, with limited signal above 10~keV.  In order to retain an 8~keV fitting band that can constrain the ionizing continuum, we set an upper energy bound of 10.4~keV.  The full spectral model was modified by line-of-sight absorption in the Milky Way via the ``hot'' model, with the column density allowed to vary.  The continuum flux components and photoionization within the system are detailed below.  The full model is detailed in Table 1, with the zones ordered to indicate the sequence in which they see radiation from the central engine.  The full spectra and best-fit models are shown in Figure \ref{fig:best_broad}.

We find that a single disk blackbody component does not fully capture the continuum flux, especially toward 10.4~keV.  The addition of a power-law component improves the fit, but diverges at low energy and distorts photoionization calculations.  We therefore adopted a model consisting of a disk blackbody (``dbb'' in SPEX), and thermal Comptonization (``compt'' in SPEX).  The parameters of the ``dbb'' model include the temperature at the point of peak emissivity and a flux normalization.  The parameters of the ``compt'' model include the seed temperature (fixed to the ``dbb'' temperature in our fits), the electron temperature, the optical depth of the electron cloud, and a flux normalization.  We found that even when the disk and seed temperatures are coupled, the parameters of this model are not well constrained, but a very low electron temperature and small optical depth are always preferred.  We therefore fixed representative values of $kT_{e} = 1$~keV and $\tau = 0.001$ in all fits.  

This simple but physical model allows for an excellent characterization of the continuum across the 2.4--10.4~keV band.  The spectra are well described when the disk and electron temperatures are linked between the ``steady'' and ``dip'' flux phases, with only their normalizations are allowed to vary.  This indicates that the shape of the ionizing continuum varies minimally between the phases. The continuum shape partially determines the radius at which thermal winds may be launched \citep{begelman1983}, so this radius is likely to also vary minimally between the flux phases.  

\subsubsection{Strong Fe XXVI and XXV absorption lines}

We proceeded to add photoionization zones using the ``pion'' model.  For simplicity, solar abundances were assumed; future work will examine potential abundance enhancements.  Following work by \citet{miller2015} demonstrating the importance of re-emission even in absorption--dominated wind spectra, we defined zones as consisting of absorption and re-emission.  Within a given zone, the column density ($N_\mathrm{H}$), ionization parameter (log $\xi$), and turbulent velocity ($\sigma_\mathrm{abs}$) were coupled between the absorption and re-emission spectra, while the bulk velocity shifts ($v$) were allowed to vary.  Geometric covering factors for the absorption ($f_{cov}$) and emission ($\Omega$, normalized by $4\pi$) were allowed to vary independently since the geometry is not known a priori. Where required, additional broadening of the emission component was modeled by convolving it with a Gaussian, leading to an independent $\sigma_\mathrm{emi}$.

\begin{figure*}
 \begin{center}
  \includegraphics[width=1.0\columnwidth]{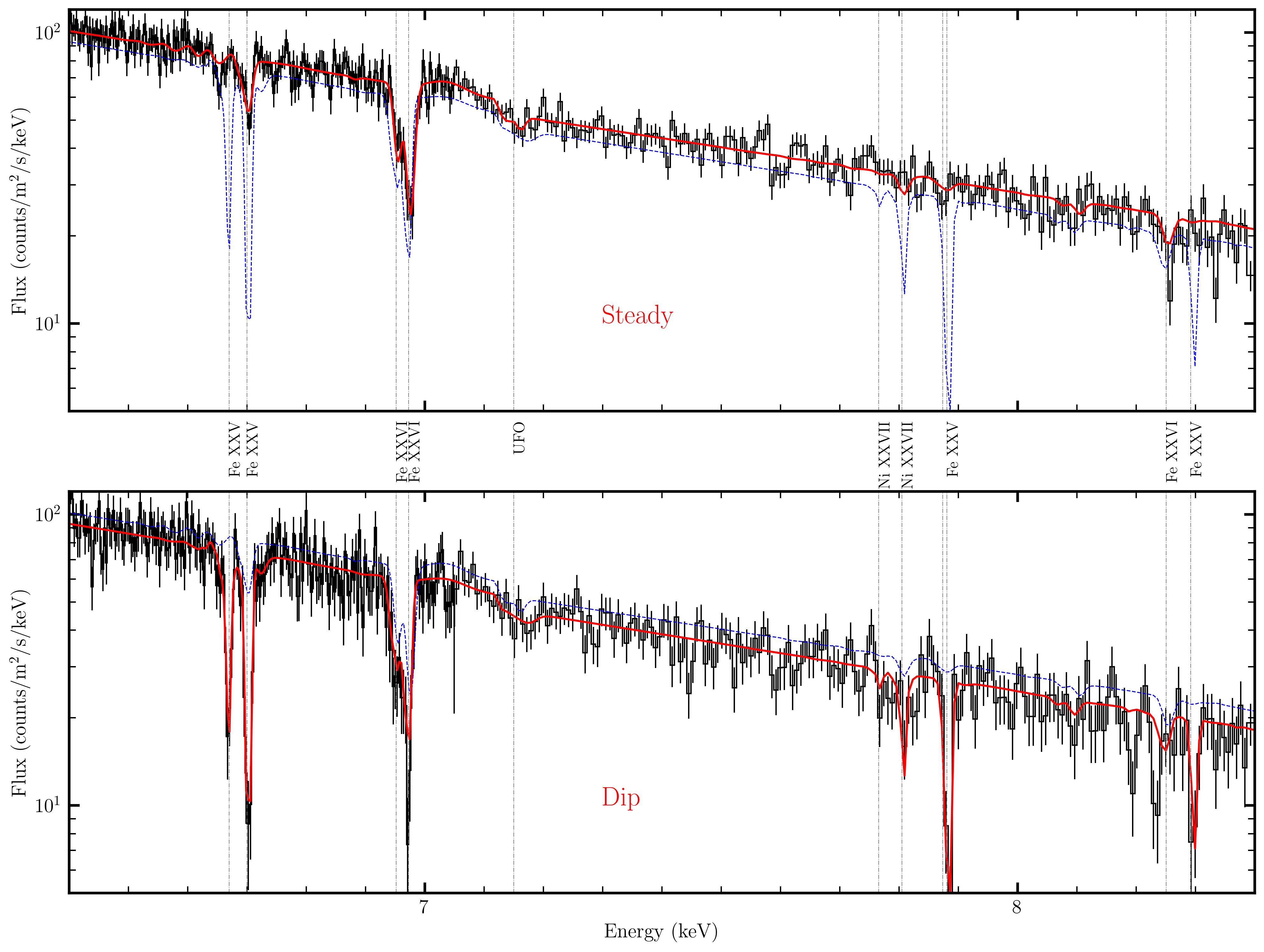} 
 \end{center}
\caption{Resolve spectra of 4U~1630$-$472 in the Fe~K band and separated by flux interval.  In both phases, the best-fit ``pion'' photoionization model is shown in red; it includes four independent absorption and re-emission zones.  In each panel, the best-fit to the complementary phase is indicated with a dashed blue line.  Line identifications are indicated between the panels, with dashed gray lines indicating the expected rest-frame laboratory energy.  The low flux phase includes a zone with strong Fe XXV He-$\alpha$, -$\beta$, and -$\gamma$ lines (the mid-$\xi$ zone in Table 1).  Note that the H-like Fe XXVI complex varies little between the phases (the high-$\xi$ zone in Table 1).  The feature at 7.15~keV likely represents Fe~XXVI absorption shifted by $v = -0.027c$ (the UFO zone in Table 1); dust absorption and other alternatives are statistically disfavored.  The spectra were binned using the ``optimal binning'' algorithm but additional binning is adopted above 7~keV in this figure for visual clarity.  Please see the text for additional details.}
\label{fig:best_fek}
\end{figure*}

\begin{figure*}
 \begin{center}
  \includegraphics[width=1.0\columnwidth]{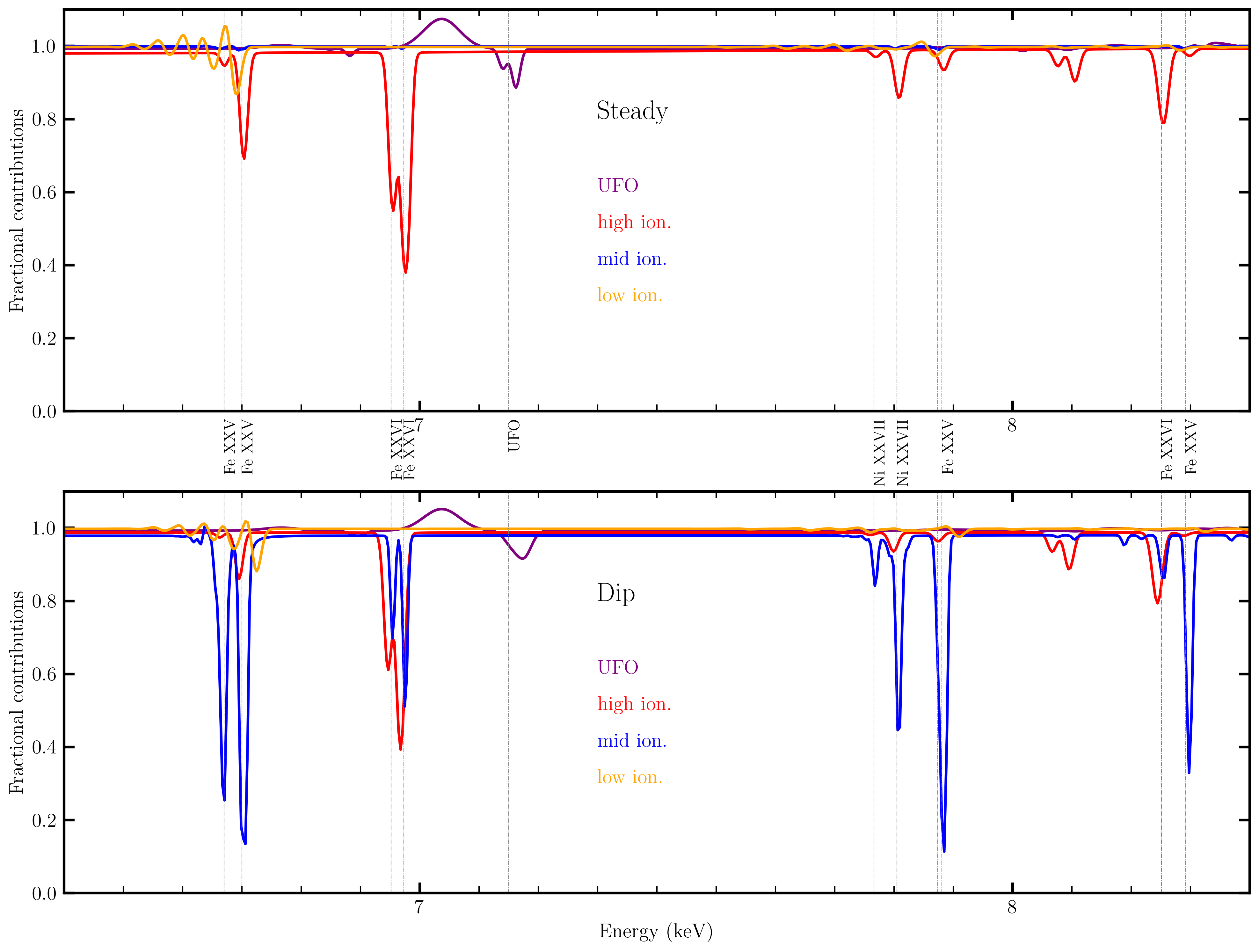} 
 \end{center}
\caption{The fractional importance of different photoionization zones in our best-fit model, relative to the local continuum flux.  The zones correspond to those detailed in Table 1.  Small vertical offsets were enforced to improve clarity in crowded regions.  The strong variation in the ``mid-$\xi$'' absorber is also evident in the prior figure, and may signal that an axially asymmetric geometry passed into the line of sight.}
\label{fig:model_ratio}
\end{figure*}

\begin{figure*}
 \begin{center}
  \includegraphics[width=1.0\columnwidth]{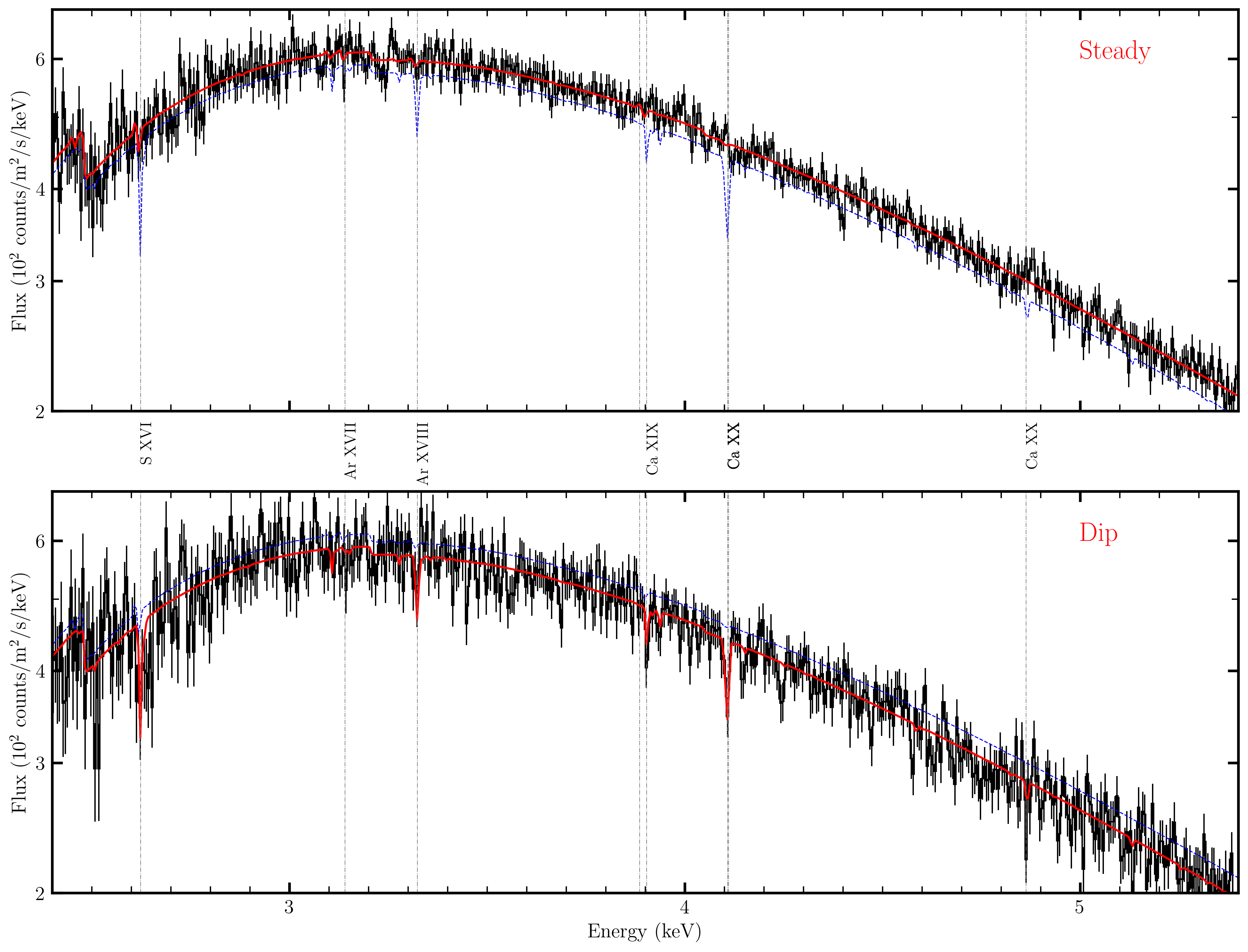} 
 \end{center}
\caption{Resolve spectra of 4U~1630$-$472 in a low-energy band and separated by flux interval.  In both flux phases, the best-fit ``pion'' photoionization model is shown in red; it includes four independent absorption and re-emission zones.  In each panel, the best-fit to the complementary phase is indicated with a dashed blue line.  Line identifications are indicated between the panels, with dashed gray lines indicating the expected rest-frame laboratory energy.  The spectra were binned using the ``optimal binning'' algorithm but additional binning is adopted in this figure for visual clarity.  Please see the text for additional details.}
\label{fig:best_low}
\end{figure*}

First, we focus on the strongest absorption lines. Figures \ref{fig:best_broad} and \ref{fig:best_fek} show that strong H-like Fe~XXVI absorption lines are common between the steady flux and dip phases, but that He-like Fe XXV absorption is {\em much} stronger in the dip phase.  Initial fits found that no single-zone model could simultaneously fit the observed Fe XXV and Fe XXVI line complexes.  Including two zones, with the continuum first passing through the highly ionized gas producing Fe XXVI absorption (the ``high-$\xi$'' zone in Table 1) and then less ionized gas mainly producing Fe XXV absorption (the ``mid-$\xi$'' zone in Table 1), adequately describes these features (see Figure \ref{fig:model_ratio} to understand the model in more detail), and therefore the bulk of the observed absorption.  The ``mid-$\xi$'' zone also produces strong absorption lines in the softer band, such as S, Ar, and Ca (Figure \ref{fig:best_low}). These lines are only seen in the ``dip'' phase.

The behavior of 4U~1630$-$472 observed with XRISM makes it unlikely that the bulk of an outflow could suddenly be launched, halted, or significantly altered.  The small change in source luminosity is unlikely to have a dramatic impact on a radiatively driven wind.  That modest change and the negligible change in spectral shape make it unlikely that the Compton radius changed, altering the state of a thermal wind.  Modest changes in luminosity reflect a correspondingly modest change in mass accretion rate and the magnetic viscosity that could launch an MHD wind, again making it unlikely that a magnetic wind would sharply change character.  For these reasons, the dramatic increase in mid-$\xi$ zone in the dip phase is unlikely to reflect a marked change in a wind and its column density.  We therefore choose to capture changes in the mid-$\xi$ zone and other absorbers in terms of covering factor rather than column density.

\begin{figure*}
 \begin{center}
  \includegraphics[width=0.7\columnwidth]{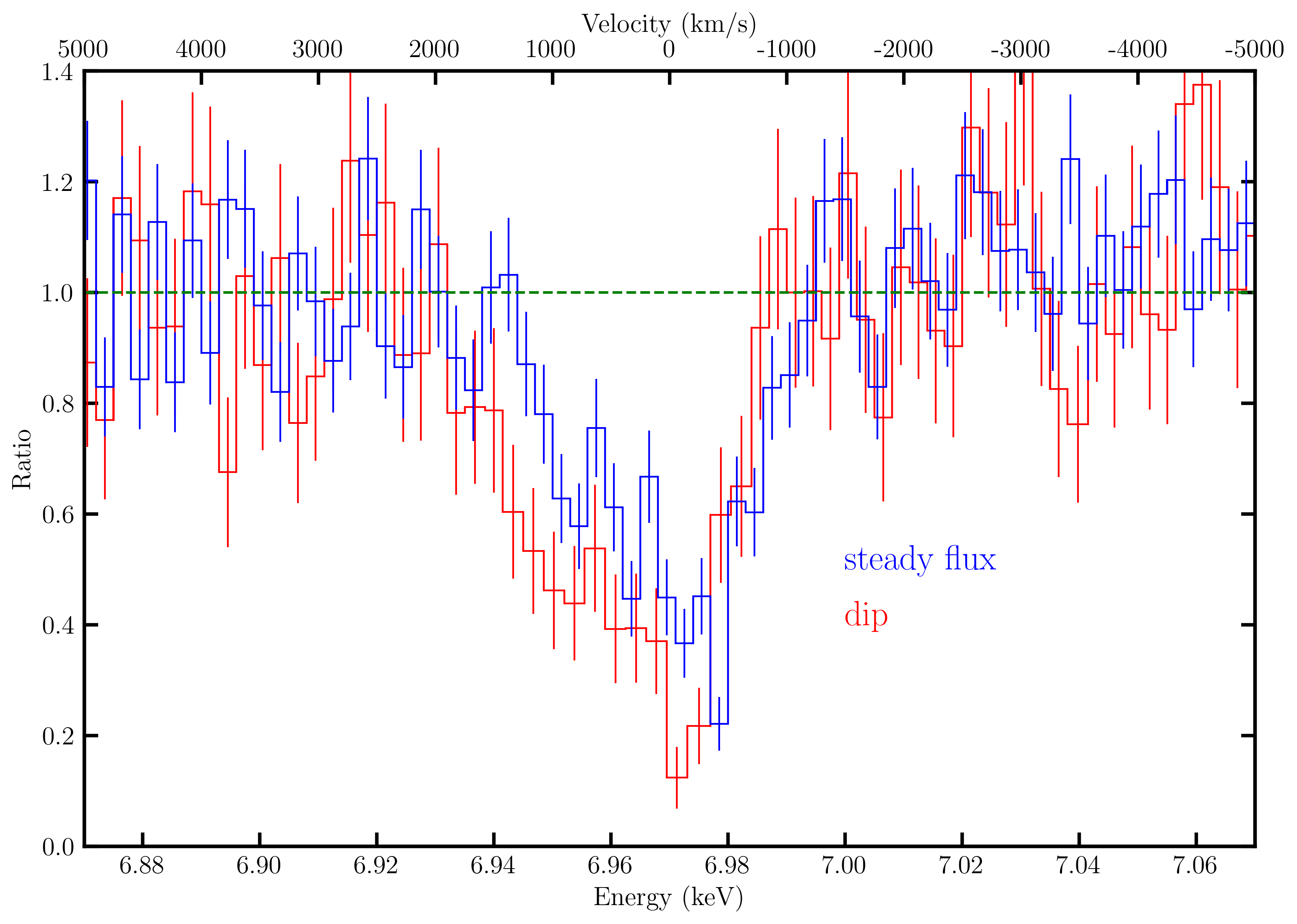} 
 \end{center}
\caption{The ratio of the ``steady'' flux and ``dip'' intervals to the continuum, without modeling the absorption.  In this model-independent representation, the velocity shift in the ``high-$\xi$'' zone (dominated by H-like Fe XXVI, a doublet centered at 6.97~keV) is clear.  Our best-fit model measures a small but significant blue-shift in the ``steady flux'' phase, $v = -130\pm 30~{\rm km}~{\rm s}^{-1}$, and a small but significant red-shift in the ``dip'' phase, $v = 220\pm 30~{\rm km}~{\rm s}^{-1}$ (see Table 1).  The data are shown with ``optimal'' binning.  }
\label{fig:hixi_vel}
\end{figure*}

Without fitting models, it is apparent that the velocity of the Fe~XXVI line(s) changes between the flux phases (Figure \ref{fig:hixi_vel}).  The line is initially slightly blueshifted in the ``steady'' phase ($-130\pm30$~km~s$^{-1}$), but significantly redshifted in the ``dip'' phase ($+220\pm30$~km~s$^{-1}$).   This indicates that the ``high-$\xi$'' absorber may also be impacted by the potentially geometric changes inferred in this observation, despite is proximity to the central engine.

\subsubsection{Other features}
Next, we examined fainter but statistically significant absorption features.
In the steady flux phase, absorption just to the {\em red} of the Fe XXV resonance line at 6.700~keV is not fit by the corresponding intercombination line at 6.670~keV.  Similarly, absorption to the {\em blue} of the Fe XXV resonance line is evident in the dip phase.  A lower-ionization zone that changes strongly in velocity ($v = 490\pm 150~{\rm km}~{\rm s}^{-1}$ to $v = -1000^{-150}_{+550}~{\rm km}~{\rm s}^{-1}$) provides an simple description of these line features (the ``low-$\xi$'' zone in Table 1).  This characterization may not be unique, but the additional screening provided by the mid-$\xi$ may help to lower the ionization of more distant gas that is then accelerated and seen as the low-$\xi$ zone.  Calculating the Akaike Information Criterion as implemented by \citet{emman16}, the inclusion of this zone improves the metric by $\Delta{\rm AIC} = 6$, indicating that its inclusion is significant (changes of $\Delta{\rm AIC} < 2$ are not significant).

Finally, broad absorption features centered close to 7.15~keV are evident in both phases, and can be fit as fast ($v \simeq 0.027-0.033c$), variable, highly ionized outflow that may carry a significant mass flux and kinetic power.  The combination of its high ionization and large velocity signal that this putative wind component likely originates closest to the central engine (see below), so it was fit as the first absorbing layer and is listed as the ``UFO zone'' within Table 1.  The addition of this component improves the AIC by $\Delta{\rm AIC} = 10$, indicating that it is also significant.   (The covering factor of the coupled emission exceeds unity, likely indicating that the column density of the emitting and absorbing gas is not the same; the covering factor effectively acts as a flux normalization.)

With the inclusion of these secondary photoionization zones, an excellent overall fit is achieved: $C = 5424.3$ for $\nu = 5384$ degrees of freedom (see Table 1).  This far exceeds the quality of comparably complex fits to prior Chandra spectra of winds in stellar-mass black holes \citep{miller2015}.  However, the potential importance of the UFO zone requires additional tests:  

First, we investigated whether or not the feature could be caused by dust.  An exhaustive set of tests is beyond the scope of this paper, but olivine is known to produce a weak, sharp feature at 7.13~keV \citep{rogantini2018}.  The fast outflow was removed from the total model, and amorphous olivine was allowed to act on the continuum via an ``amol'' component within SPEX.  Whereas the UFO zone improved the overall fit statistic by $\Delta C = 23$, olivine only improved the fit by $\Delta C = 12$. This is unsurprising since the predicted feature is weak relative to the K edge, sharp, and carries an asymmetric profile that differs from an absorption line \citep{rogantini2018}.  

Second, we investigated whether or not K$\beta$ absorption lines from charge states with filled 2p shells, Fe X-XVII, could describe this putative fast outflow feature.  In these tests, the fast outflow was removed from the model, and a pion absorber was included with a bounded ionization parameter to ensure Fe X-XVII line production.  In all such fits, sufficiently strong K$\beta$ line absorption is accompanied by a far stronger edge that is entirely inconsistent with the data.  No model of this kind provided any improvement to the fit statistic, and this explanation is also disfavored. 

Third, an emission line contribution from ionized absorber out of the line-of-sight can be considered. When the fit is performed without the UFO component, a positive residual remains around 6.95–-7.10 keV, as well as the 7.15~keV dip (see the UFO line in Figure \ref{fig:model_ratio}). In the UFO scenario, this positive residual is a P~Cygni profile of the UFO gas. Alternatively, if the ``high-$\xi$'' component, which produces the strong Fe XXVI absorption structure, is located around several hundred $R_\mathrm{g}$, it can produce an emission line broadened by several tens of eV due to Keplerian motion (see Figure 5 in \citealt{trueba2022}). This scenario will be discussed in detail in a separate paper.

\subsection{Characteristics of Photoionized Plasma Components and Their Possible Origins}

Table 1 details our best-fit photoionization model.  Using the ionization parameter formalism, line-of-sight bulk velocities, variability, and broadened re-emission, basic constraints on the absorption radius of each zone can be derived (essentially an upper limit on the launching radius, if the zone has a bulk velocity).  Then, it is possible to identify which zones are likely to escape to infinity or return to the accretion flow as ``failed winds,'' and to speculate on the mechanisms that might lift gas into our line of sight and/or drive some zones out of the system.  

One crucial radius to consider is the Compton radius, at which heating from the central engine can raise the temperature of the disk surface to the local escape speed.  Following \citet{begelman1983}, $R_\mathrm{C} = 1.0\times 10^{10}~(M_\mathrm{BH}/M_{\odot}) T_{\mathrm{C},8}^{-1}$, and approximating the Compton temperature with the disk temperature of $kT = 1.9$~keV or $T = 2.2\times 10^{7}$~K, $R_\mathrm{C} \simeq 4.5\times 10^{11}$~cm for $M_\mathrm{BH} = 10~M_{\odot}$.  In gravitational radii, this is $R_\mathrm{C} = 3.0\times 10^{5}~GM/c^{2}$.  \citet{woods1996} notes that winds may be driven from $0.2~R_\mathrm{C}$, or $6\times 10^{4}~GM/c^{2}$.

The ionizing {\em luminosity} in the steady flux and dip phases is measured to be $L_\mathrm{ion} = 6.00\pm 0.02 \times 10^{37}~{\rm erg}~{\rm s}^{-1}$ and $L_\mathrm{ion} = 6.17\pm 0.02 \times 10^{37}~{\rm erg}~{\rm s}^{-1}$.  The drop in flux in the dip phase is likely the result of increased internal obscuration, {\em not} a drop in continuum luminosity. Note that these luminosity values correspond to Eddington fractions of $\lambda = 0.046$ ad $\lambda = 0.047$, respectively, for $M_\mathrm{BH} = 10~M_{\odot}$.  

The ionization parameter of the high-$\xi$ zone implies $r \leq 1.4\times 10^{9}~{\rm cm}$ or $r \leq 850~(M/10~M_{\odot})~GM/c^{2}$.  The high-$\xi$ zone is nominally blueshifted initially and then slightly but significantly redshifted during the subsequent dip phase.  The escape speed from this radius is $v \simeq 0.05c$, making it unlikely that the gas can escape even if it is launched vertically, and even if it is partly rotationally supported.  Given the small inferred launching radius, the high-$\xi$ zone may be a failed magnetic wind, and/or associated with a static but turbulent disk atmosphere \citep{trueba2022}.  The close correspondence between the velocity broadening expected from the plasma temperature ($kT = 0.49$~keV, giving $\sigma \sim 200~{\rm km}~{\rm s}^{-1}$) and the measured broadening ($\sigma = 250\pm 20~{\rm km}~{\rm s}^{1}$) may also suggest that the high-$\xi$ component is confined to a narrow run of radius.  

A steady, axially symmetric disk atmosphere should likely have consistent turbulent and bulk velocities across relatively modest changes in luminosity.  It is possible that the high-$\xi$ component could be described in this way, and that its apparent velocity shift may really be variability better ascribed to other absorbers.  To test this possibility, we linked the bulk velocity in the high-$\xi$ component between the steady flux and dip phases, and re-fit the data.  The overall fit is worse by $\Delta C = 33$ for one degree of freedom, indicating that this possibility is disfavored by the data at the $5\sigma$ level of confidence.  This may simply indicate that the disk atmosphere is not homogeneous, with radial and vertical structure, and an interaction with the dip that is not yet clear.

The mid-$\xi$ zone shows the most dramatic variation between the steady flux and dip phases.  The strength of the Fe XXV absorption lines increases by an order of magnitude.  Ordinarily, a drop in luminosity and ionization parameter would make Fe XXV lines stronger relative to Fe XXVI lines, but in 4U~1630$-$472 the ionizing luminosity slightly {\em increased} during the dip phase.  Given that the change in luminosity is relatively small and the change in spectral shape is negligible, it is unlikely that a dramatic change arose that led to the sudden production of a thermal wind. 
The absence of a thermal wind is consistent with what the theory predicts. In the framework of Compton-heated winds, we expect a critical luminosity below which the plasma cannot be launched as a wind (e.g., \citealt{diaztrigo2013}). The luminosity at the XRISM epoch ($\sim 6 \times 10^{37}$ erg s$^{-1}$ or $\lambda \sim 0.05~ (10~M_{\odot}/M_{BH})$) is below the critical luminosity. 

The mid-$\xi$ zone may instead be associated with an axially asymmetric structure.  \citet{armitage1998} show that at low accretion rates, the region where the impact stream hits the outer disk can cool efficiently, allowing the stream to overflow the disk until the two connect at the point of closest passage.  This axially asymmetric structure would give rise to a long dip of the sort that is observed.  The local disk atmosphere above this geometry could be consistent with the mid-$\xi$ zone.  Again rearranging the ionization parameter formalism, we find that the mid-$\xi$ zone originates at $r \leq 1.1\times 10^{11}~{\rm cm}$ from the black hole, or $r \leq 7.3\times 10^{4}~(M/10~M_{\odot})~GM/c^{2}$.  Longer dips with higher ionization parameters than the normal absorption dips have been observed in some other X-ray binaries (e.g., \citealt{shidatsu2013}), and may arise through similar geometries.   Similar to the high-$\xi$ absorber, the close correspondence between the anticipated thermal broadening versus measured broadening for the mid-$\xi$ component ($kT = 0.11$~keV, giving $\sigma = 90~{\rm km}~{\rm s}^{-1}$, versus $\sigma = 120\pm 10~{\rm km}~{\rm s}^{-1}$) may suggest that it is not accelerated over a run of radii, but relatively fixed within the binary system.

The low-$\xi$ zone appears to be most strongly affected by the modest change in luminosity between the initial and later parts of the observation, and/or the putative geometric changes leading to the obscuration captured by the mid-$\xi$ absorber.  The low-$\xi$ component is initially red-shifted by $v = 490\pm 150~{\rm km}~{\rm s}^{-1}$, but blue-shifted by $v = -1000^{-1200}_{+550}~{\rm km}~{\rm s}^{-1}$ during the dip phase.  It is possible that this component sees a drop in flux if the mid-$\xi$ zone does indeed block some flux from the central engine; with an ionization of ${\rm log}\xi = 2.96\pm 0.08$, the low-$\xi$ zone is on the edge of the point at which radiation pressure on lines may be effective.  Using the ionization parameter formalism and again assuming unity filling factors, the low-$\xi$ zone nominally originates at $r \leq 2.9\times 10^{13}~{\rm cm}$ or $r \leq 1.9\times 10^{7}~(M/10~M_{\odot})~GM/c^{2}$.  This corresponds to the extreme outer disk, if it is even within the binary system. Unlike the mid- and high-$\xi$ components, the measured broadening of the low-$\xi$ component is greater than that predicted by thermal broadening ($\sigma = 350^{+70}_{-60}~{\rm km}~{\rm s}^{-1}$, versus $\sigma = 70~{\rm km}~{\rm s}^{-1}$ for $kT = 0.057$~keV).  This may suggest radiative acceleration over a range in radius.

The UFO zone nominally has a lower ionization parameter than high-$\xi$ zone, but it carries a much higher blue-shift.  At ionization parameters above ${\rm log}\xi = 3$, radiation pressure on lines is not able to drive winds to high speeds, so it is possible that the observed speed partially reflects the escape speed (modulo a projection factor) at the point of origin.  For this reason, the UFO zone may have originated interior to the high-$\xi$ zone.  The blue-shift of $v = -8600\pm 1000~{\rm km}~{\rm s}^{-1}$ measured in the dip phase implies $r \leq 2400~GM/c^{2}$ (where the limit reflects that any de-projection leads to a smaller launching radius).  Rearranging the ionization parameter formalism, $r \leq L/N_{H}\xi$ (the upper limit obtains because this assumes geometric and gas filling factors of unity), we find that the UFO zone likely originates at $r \leq 3\times 10^{10}~{\rm cm}$ from the black hole, or $r \leq 2.0\times 10^{4}~(M/10~M_{\odot})~GM/c^{2}$.  The disparity likely reflects that the volume and gas filling factors are far less than unity.  Assuming that the gas in the UFO zone originates at radii below $r \leq 2400~GM/c^{2}$, it is launched an order of magnitude below the smallest radius at which thermal winds are possible.  Moreover, even its projected speed greatly exceeds the $v = 200~{\rm km}~{\rm s}^{-1}$ maximum found in simulations \citep{higginbottom2017}.  As with the low-$\xi$ zone, the higher measured velocity broadening ($\sigma = 500~{\rm km}~{\rm s}^{-1}$) relative to the thermal broadening predicted by pion (the plasma temperature of $kT = 0.50$~keV predicts $\sigma = 200~{\rm km}~{\rm s}^{-1}$) may suggest acceleration over a run of radii.  The UFO zone may represent a magnetic wind from the inner disk.

\subsection{Uncertainties in velocity measurements}

To understand the shifts and the velocity dispersion of the ionized absorbers, we need to estimate the contributions of the binary system itself.  The heavy extinction of the source has hampered direct radial velocity measurements of 4U~1630$-$472 via optical spectroscopy.  However, we can place some constraints on the orbital velocity of the black hole and the systemic velocity using proxies. The fact that 4U~1630$-$472 shows outbursts frequently suggests that it has a large disk, and that the mass transfer rate from the companion star may be close to the threshold for persistent accretion \citep{coriat2012}.  Following the method in \citet{coriat2012}, we estimate the integrated X-ray luminosity in a normal outburst (such as the one around MJD 57000) divided by the outburst recurrence time ($\sim 600$ days) of $L \sim 10^{38}$ erg s$^{-1}$, using the 2--20 keV MAXI light curve and assuming a distance of 11.5 kpc. From this value,  we estimate the mass transfer rate of $\dot{M} \sim 10^{18}$ g s$^{-1}$.  Comparing the relation of the orbital period and the mass transfer rate given in \citet{coriat2012}, we obtain the orbital period of $P \gtrsim 50$ hours. Applying this period to the binary mass function, we obtain the radial velocity of the black hole of $\sim 90$ km s$^{-1}$, assuming a relatively large companion mass of $M_{\rm c} = 3 M_\odot$ and a black hole mass of $M_{\rm BH} = 10 M_\odot$, respectively. The value roughly scales with $M_{\rm c}$ and $P^{-1/3}$, so a smaller $M_{\rm c}$ and/or a longer $P$ give a smaller velocity.  

We next consider the galactic rotation and the natal kick velocity.
The systemic radial velocity measurements of 12 galactic black hole X-ray binaries are summarized in \citet{atri2019}. The velocity reflects both the galactic rotation and the natal kick at the birth of the black hole.  According to the list, all the systemic velocities fall within $\pm 100$  km s$^{-1}$ except for one source. Thus we consider that it is reasonable to assume the systemic radial velocity of 4U~1630$-$472 to be less than 100 km s$^{-1}$. For a cross-check, we estimate the radial velocity of 4U~1630$-$472 due to the Galactic rotation. If we assume that the Galactic rotation of the solar system is the same as the local standard of rest and 4U~1630$-$472 has a circular orbit around the Galactic center, the radial velocity of 4U~1630$-$472 relative to the solar system is estimated to be $\sim$ 60 km s$^{-1}$. This falls within the range of $ < 100$ km s$^{-1}$ and is consistent with the above assumption.

For completeness, we also note the velocity of XRISM, and velocity shifts due to Earth's motion within its orbit.  The velocity of XRISM satellite with respect to  the Earth is $\sim 8$ km s$^{-1}$, so its contribution to the line-of-sight velocities of is negligible. The line-of-sight velocity due Earth's orbital motion around the barycenter of the solar system is $\sim 26$ km s$^{-1}$.  Its variation is $\lesssim 0.2$ km s$^{-1}$ in this observation, which is small compared with the uncertainties in the velocities that we measured in the spectral fitting. 

Combining all these estimates, the uncertainty in the intrinsic radial velocity of the ionized absorbers is $\lesssim 220$ km s$^{-1}$. This is comparable to the velocities of the absorption lines for the high-$\xi$ zone and the mid-$\xi$ zone, and hence whether they are outflowing or inflowing is uncertain (regardless, they may be bound, and unable to completely escape).  The change in the velocities of motions of the binary system and the observer velocity during the observation depends mostly on that of the binary orbital motion and would be $\lesssim 180$ km s$^{-1}$ if the orbital period is relatively short.  

\section{Discussion}
We have analyzed complex, variable absorption in the Resolve spectrum of 4U~1630$-$472. Whereas most prior studies of ionized absorption in stellar-mass black holes can be understood in terms of blue-shifted winds from the accretion disk, this high-resolution spectrum of 4U~1630$-$472 reveals different aspects of the accretion flow and its geometry.  In this section, we examine the role of Eddington fraction and ionizing luminosity in creating these differences, and explore the flows and geometries that are inferred through our models.  We also remark on the limitations of our analysis, and comment on future prospects for studies of stellar-mass black holes with XRISM.

\subsection{Absorption spectra and Eddington fraction}
We observed 4U~1630$-$472 at an Eddington fraction of $\lambda \sim 0.05~ (10~M_{\odot}/M_{BH})$, which is one of the lowest fractions at which absorption lines have been detected (see \citealt{ponti2012}, \citealt{parra2024}).  Whether or not velocities within the binary system would render the small velocity shifts that we have measured in the primary zones as redshifts or blueshifts, the associated gas is unlikely to escape the system.  (If the very fast outflow is indeed a disk wind, it may escape the system, and drive significant feedback; see below).  In light of our results, it may be helpful to consider a few simple questions: (1) Is there an Eddington fraction below which escaping winds cannot be driven?  (2) Is there an Eddington fraction below which even failed winds are not launched?  (3) Is Eddington fraction the only important factor in driving winds (escaping or failed)?  A complete answer to these questions is beyond the scope of this paper, but we can examine past observations of 4U~1630$-$472.

A prior Chandra/HETG spectrum of 4U~1630$-$472 at $\lambda_{Edd} \sim 0.06$ ruled out the presence of ionized absorption lines at high statistical confidence \citep{gatuzz2019}.  In part, this reflects that Eddington fraction and spectral hardness are coupled in complex ways in stellar-mass black holes; the observation in question required a $\Gamma = 2.1$ power-law component in addition to a thermal disk component.  A power-law like this is excluded in our Resolve spectrum.  It is possible, then, that a much higher ionization made a wind difficult to detect (e.g., \citealt{diaztrigo2014}), and that prior observations did not explore the full range of relevant Eddington fractions.

Different outbursts from the same stellar-mass black hole can be very different (see, e.g., Figure 1).  Some outbursts display a full range of states, whereas others might remain in the ``low/hard'' state.  Only considering those that show a variety of states, transitions between one state and another may not happen at the exactly the same apparent Eddington fraction (e.g., \citealt{capitanio2015}; for a review, see \citealt{rm2006}).  Our observation demonstrates the importance of not expecting every outburst to be the same.  It also demonstrates that failed winds and potentially very fast winds are not restricted to high Eddington fractions, but possible at $\lambda_{Edd} \sim 0.05$.  Future XRISM programs that monitor soft, disk--dominated states that persist to even lower Eddington fractions may resolve the role of Eddington fraction in driving different winds, and separate the roles of Eddington fraction and hardness in this matter.

Simulations of black hole disks and winds reveal significant variability in all wind parameters, over radius, scale height, and time (e.g., \citealt{vourellis2019}).  In principle, a single line of sight intercepting different parts of a relatively continuous flow may appear as distinct zones.  We cannot exclude this scenario for the absorption that Resolve has detected in 4U~1630$-$472.  However, we have detected zones with different and variable velocity shifts that nominally require different driving mechanisms, and at least one zone is potentially tied to a broad disk structure rather than an outflow (see below).  On balance, then, it is more likely that the observed absorption is not part of a single outflow.  In contrast, the disk winds detected in 4U~1630$-$472 at higher Eddington fractions consistently show blue-shifts in every zone, and their AMD is consistent with a single flow \citep{trueba2019}.

\subsection{Axially asymmetric structure in the accretion flow}
The flux change shown in Figure 3 is very clear owing to the sensitivity of Resolve, but it only represents a 10\% drop in apparent flux.  As noted above, this is likely incompatible with a large change in the radius at which a thermal wind could be launched, and likely also incompatible with such a wind suddenly turning on or off.  
The observed flux trend is both longer in duration and smaller in fractional depth than typical ``dips'' lasting $\leq 100$~s \citep{kuulkers1998}.  Moreover, every zone in the ionized absorption spectrum is affected by the small flux change seen in Figure 3; it is not merely the case that outer zones are affected.  A large-scale geometric change is required to achieve this.  

Warped, precessing disks can reveal different parts of a wind over time \citep{kosec2023}, but the super-orbital time scales are likely too long to describe the phenomenon observed in 4U 1630$-$472.  Rather, an accretion stream passing into our line of sight because it has overflowed the outer disk may explain the extended flux dip and mid-$\xi$ zone in 4U 1630$-$472.
The resulting axially asymmetric geometry may extend to relatively small radii, hitting the disk at the point of closest passage. 
\citet{armitage1998} show that this scenario only manifests at relatively low accretion rates.  This Eddington requirement and poor sampling of an unknown orbital period may explain why dips of this sort have not previously been detected in long observations of 4U 1630$-$472 with other X-ray telescopes.  Long dips in other X-ray binaries (see, e.g., \citealt{shidatsu2013}) may also be explained in this way. 

\subsection{Failed winds and disk atmospheres}
The two primary absorption components that we have observed in 4U~1630$-$472 have very low velocities, and potentially redshifts.  Even for the largest possible radii that are obtained when filling factors of unity are assumed, the flows are unlikely to escape from the system.  This is true even if the gas is launched almost vertically from the disk.  Given that the velocities and covering factors of these components change, they may represent ``failed winds'' that circulate within the system but never escape to distribute mass and kinetic power into the broader environment.  These components could also be associated with the disk atmosphere (e.g., \citealt{rozanska2014}), but given that both successful and failed winds must be driven from the disk, this may only be a matter of terminology.

The failed wind or turbulent atmosphere could be produced through either X-ray irradiation or magnetic processes. X-ray irradiation on the disk surface can produce  a hot plasma that is gravitationally bound to the system. This could explain the observed very low velocity, although how much turbulent velocity is expected in this process and whether or not it can explain the velocity variation are unclear. It is also possible that the observed component is the failed wind formed in magnetic processes. 

In order to explain differences between theoretical predictions of disk properties and observations of novae, \citet{nixon2020} describe strong surface flows along the disk that could shift a large fraction of the accreting mass without it escaping.  In this scenario, the surface flows are driven by magnetic stresses.  Distributed surface flows of this kind may also be implied by recent X-ray polarimetry of 4U~1630$-$472, wherein an unexpectedly high polarization fraction of 6--10\% is detected \citep{ratheesh2024}.  The flow implied by polarimetry may need to be mildly relativistic, and potentially similar to the tentative very fast outflows that we have modeled in the Resolve spectra.  Future observations of stellar-mass black holes that pair XRISM and IXPE -- and particularly observations of 4U~1630$-$472 -- may be able to test these and other models in detail.

\subsection{The fast wind component}
Rewriting the basic equation of the mass outflow rate, $\dot{M} = \Omega \mu m_{p} n r^{2} v$ using the ionization parameter formalism, $\dot{M} = \Omega \mu m_{p} (L/\xi) v$.  Here, $\Omega$ is the fraction of the $4\pi$ solid angle that is covered; $\mu$ is the mean atomic weight, assumed to be 1.2; and $m_{p}$ is the mass of the proton.  Using the values in Table 1 and assuming $\Omega = 0.5$, mass outflow rates between $\dot{M} = 0.47-3.0\times 10^{17}~{\rm g}~{\rm s}^{-1}$, or $\dot{M} = 0.74-4.7\times 10^{-9}~M_{\odot}~{\rm yr}^{-1}$ are implied for the tentative fast wind component.  This formulation of the outflow rate accounts for the geometric filling factor of the wind zone relative to the total volume, but assumes a filling factor of unity within the wind zone.  

These outflow rates can be contrasted with the implied mass accretion rate, $\dot{M}_{acc} = L / \eta c^{2}$, or $\dot{M} = 1.1\times 10^{-8}~M_{\odot}~{\rm yr}^{-1}$ for an assumed efficiency of $\eta = 0.1$.  It is not the case that more gas is being expelled than accreted, but a significant fraction of the gas that starts in the outer disk may be lost in the UFO-like wind.  Taking the kinetic luminosity to be $L_{kin} = \frac{1}{2} \dot{M} v^{2}$, the power in this wind component is $L_{kin} = 0.14-1.4\times 10^{35}~{\rm erg}~{\rm s}^{-1}$, or 0.02-0.2\% of the radiative power.  In this sense, the UFO-like wind may remove a significant fraction of the gas available for accretion onto the black hole, but falls just below the threshold for ``fierce feedback'' in massive black holes.  

At the opposite end of the mass scale, the recent discovery of a UFO in Mrk 817 at an Eddington fraction of $\lambda_{Edd} = 0.008-0.016$ \citep{zak2024} contrasts with UFOs seen in Eddington-limited sources like PDS 456 \citep{nardini2015}.   The detection of a fast wind in a stellar-mass black hole at just $\lambda_{Edd} = 0.05$ could signal that a common mechanism launches fast winds in sub-Eddington phases.  Magnetic pressure from viscosity within the disk is a potential candidate.  However, it is possible that different mechanisms (or, combinations of mechanisms) work to drive fast winds at opposite ends of the black hole mass scale.

The UFO in 4U~1630$-$472 appears to have varied little between the ``steady'' and ``dip'' phases (see Table 1).  Unlike the high-$\xi$ zone, velocities linked to motions within the binary are small compared to the outflow velocity of the UFO itself.  More broadly, this consistency may only reflect that the most plausible UFO driving mechanisms are not affected by very modest changes in the spectrum and mass accretion rate.  Future Resolve observations of stellar-mass black holes that detect marked changes in the central engine may help to reveal UFO driving mechanisms and characteristics.

Although the fast wind tentatively detected with Resolve likely occurs at the lowest Eddington fraction yet, fast winds have often been detected in stellar-mass black holes below the Eddington limit.  Flows with projected speeds between $v/c = 0.01-0.04$ 
have previously been observed in sub-Eddington states of 4U~1630$-$472 \citep{king2014}, and also in apparently sub-Eddington states of the X-ray binaries IGR J17091$-$3624 \citep{king2014}, IGR J17480$-$2446 \citep{miller2011}, GRO~J1655$-$40 \citep{balakrishnan2020}, and GRS~1915$+$105 \citep{miller2016}.  These winds were detected in sources exhibiting different coronal properties, ranging from a substantial power-law component to states with significant disk reflection.   The putative UFO in the Resolve spectrum of 4U~1630$-$472 has a lower ionization and column density than the UFO seen in 
MAXI J1348$-$630 \citep{chakraborty2021}, but exhibits properties similar to the UFO in IGR~J17091$-$3624 \citep{king2014}, potentially further indicating a diversity of wind properties tied to coronal flux levels and Eddington fraction.

\subsection{Limitations}
This analysis has divided the XRISM observation of 4U~1630$-$472 into just two parts.  The sensitivity of the resultant spectra suggests that the data could be divided into smaller time segments, and that the evolution of the absorption spectrum could be traced with higher fidelity.  It may also be revealing to compare 256s bins before, during, and after local maxima or minima, to search for absorption variability on very short time scales.  These investigations are beyond the scope of our current analysis, but they hold real potential.

Owing to the fact that the observed variability is modest in a fractional sense, we have assumed that the mid-$\xi$ component is not a wind that turned on during our observation, and inferred that changes in the observed spectra are primarily driven by geometric changes within the system.  Even if this assumption is broadly true, it cannot be entirely true.  An atmosphere and/or wind-launching region above the surface of the disk is likely to have vertical and radial gradients in all of its properties (e.g., \citealt{kosec2023}), and therefore covering factors and columns likely vary together, even if covering factors change more than columns.  In our analysis of the data, variations in column and covering factor appeared to be nearly entirely degenerate; however, future investigations may be able to break this degeneracy.

Although our analysis indicates that different absorption zones may be tied to independent structures, we cannot rule out the possibility that every zone is part of a single outflow that spans a range of gas properties.  For instance, our fits suggest that the column density is positively correlated with ionization ($N_{H} \propto \xi$), which is consistent with magnetically driven winds but counter to the predictions of thermal winds (see, e.g., \citealt{chakravorty2016}, \citealt{contopoulos2017}, \citealt{kazanas2012}, \citealt{fukumura2021}, \citealt{dyda2017}).  New simulations may be able to determine whether or not the degree of complexity that we have observed is possible in a single magnetically driven outflow, or other flows.

Finally, although we have found alternative explanations for the UFO-like wind zone to be statistically disfavored, we caution that our interpretation of the features close to 7.15~keV should be regarded as preliminary.  While it is true that similar components are found in other X-ray binaries in similar phases, this does not mean that the features in 4U~1630$-$472 -- viewed along a line of sight with a much higher column -- are intrinsic.  Our investigation of dust absorption is limited to olivine, so it is necessarily incomplete.  Future work that considers different combinations of dust compounds may find that the feature can be explained through dust along the line of sight to 4U~1630$-$472, rather than atomic absorption within the system.  It is also possible that more complex combinations of absorption from a narrow range of charge states with filled inner shells (Fe X-XVII) could explain the absorption through atomic absorption with a smaller velocity shift.  We note that even when archival Chandra spectra of 4U~1630$-$472 are stacked, the presence of a similar feature cannot be statistically required nor rejected, so this matter is also one that must partly rely on future observations.

\section{Conclusions}
The first XRISM observation of a transient stellar-mass black hole caught 4U~1630$-$472 at the tail end of a soft, disk--dominated state.  This is an ideal scenario in which to explore how disk winds vary with Eddington fraction.  However, rather than the relatively simple disk winds implied in prior studies of 4U~1630$-$472 with Chandra, NuSTAR, Suzaku, and XMM-Newton at higher Eddington fractions, Resolve obtained a very complicated and highly variable ionized absorption spectrum.  The data likely reflect a combination of failed winds, anisotropy that may be consistent with an overflowing accretion stream, and potential fast outflows with high mass transfer rates.  Future observations of 4U~1630$-$472 and other black hole transients with XRISM are needed to understand how disk winds evolve with Eddington fraction, and to better understand the complexity seen in this observation.

We acknowledge the anonymous referee for comments that improved this manuscript.  We thank Daniele Rogantini and Liyi Gu for helpful discussions.  This work was supported by JSPS KAKENHI Grant Number 
JP21K13958 (M.M.), 
JP19K14762, JP23K03459 (M.S.), 
JP20H01946 (Y.U.),
JP20K04009 (Y.T.),
JP24K17105 (Y.K.),
JP24K17104 (S.O.),
NASA grants
80GSFC21M0002 (R.B.),
80NSSC20K0733 (E.B.),
STFC grant ST/T000244/1 (C.D.),
JSPS Core-to-Core Program (grant number:JPJSCCA20220002),
and the Strategic Research Center of Saitama University.
M.M.\ acknowledges support from Yamada Science Foundation.

\bibliography{main}{}
\bibliographystyle{aasjournal}

\clearpage

\clearpage

\clearpage

\clearpage

\clearpage
\begin{table}[htb]
\title{Main X-ray Observations}
\begin{center}
\begin{tabular}{llll}
Parameter Name & Steady flux phase & ~  & Dip phase \\
\hline
Disk kT (keV) & ~ & $1.90^{+0.01}_{-0.01}$ & ~ \\
Disk Norm. ($10^{-8}$) & $1.043^{+0.004}_{-0.004}$ & ~ & $1.080^{+0.008}_{-0.004}$ \\
Compt kT$_e$ (keV) & ~ & 1.0 & ~  \\
Compt $\tau$ & ~ & 0.001 & ~ \\
Compt Norm. & $7.1^{+0.3}_{-0.3}$ & ~ & $5.6^{+0.3}_{-0.3}$ \\
\hline
UFO Abs. ${\rm N}_{\rm H}$ ($10^{22}~{\rm cm}^{-2}$) & ~ & $1.0^{+0.3}_{-0.3}$ & ~ \\
UFO Abs. log $\xi$ & ~ & $5.6^{+0.4}_{-0.3}$  &~  \\
UFO Abs. $\sigma$ (km~s$^{-1}$) & $270^{+110}_{-270}$ & ~ & $500_{-10}$  \\
UFO Abs. $v$ (km~s$^{-1}$) & $-8000^{-1000}_{+200}$ & ~ & $-8600^{-1000}_{+1000}$\\
UFO Abs. $f_{cov}$ & ~ & 1.0  & ~ \\
UFO Emi. $v$ (km~s$^{-1}$) & 0.0 & ~ & 0.0 \\
UFO Emi. $\Omega$ & $4.8^{+1.8}_{-1.6}$ & ~ & $3.1^{+2.4}_{-2.2}$\\
UFO Emi. $\sigma$ (km~s$^{-1}$) & -- & ~ & $1200\pm 400$  \\
\hline
high-$\xi$ Abs. ${\rm N}_{\rm H}$ ($10^{22}~{\rm cm}^{-2}$) & ~ & $9.7^{+0.2}_{-0.2}$ & ~ \\
high-$\xi$ Abs. log $\xi$ & ~ & $5.70^{+0.03}_{-0.03}$ & ~ \\
high-$\xi$ Abs. $\sigma$ (km~s$^{-1}$) & ~ & $250^{+20}_{-20}$ & ~  \\
high-$\xi$ Abs. $v$ (km~s$^{-1}$) & $-130^{-30}_{+30}$ & ~ & $220^{+30}_{-30}$ \\
high-$\xi$ Abs. $f_{cov}$ & 1.0 & ~ & $1.0_{-0.02}$ \\
\hline
mid-$\xi$ Abs. ${\rm N}_{\rm H}$ ($10^{22}~{\rm cm}^{-2}$) & ~ & $7.8^{+0.3}_{-0.3}$ & ~ \\
mid-$\xi$ Abs. log $\xi$ & ~ & $3.93^{+0.05}_{-0.05}$ & ~ \\
mid-$\xi$ Abs. $\sigma$ (km~s$^{-1}$) & ~ & $120^{+10}_{-10}$ & ~  \\
mid-$\xi$ Abs. $v$ (km~s$^{-1}$) & ~ & $-110^{-10}_{+10}$ &  ~ \\
mid-$\xi$ Abs. $f_{cov}$ & $0.014^{+0.003}_{-0.014}$ & ~ & 1.0 \\
mid-$\xi$ Emi. $v$ (km~s$^{-1}$) & 0 & ~ & $-80^{-70}_{+110}$  \\
mid-$\xi$ Emi. $\Omega$ & $0.0048^{+0.043}_{-0.0048}$ & ~ & $0.15^{+0.05}_{-0.06}$\\
\hline
low-$\xi$ Abs. ${\rm N}_{\rm H}$ ($10^{22}~{\rm cm}^{-2}$) & ~ &  $0.28^{+0.05}_{-0.06}$ & ~ \\
low-$\xi$ Abs. log $\xi$ & ~ & $2.96^{+0.08}_{-0.08}$ & ~ \\
low-$\xi$ Abs. $\sigma$ (km~s$^{-1}$) & ~ & $350^{+70}_{-60}$ & ~  \\
low-$\xi$ Abs. $v$ (km~s$^{-1}$) & $490^{+150}_{-150}$ & ~ & $-1000^{-150}_{+550}$ \\
low-$\xi$ Abs. $f_{cov}$ & ~ & 1.0 & ~  \\
low-$\xi$ Emi. $v$ (km~s$^{-1}$) & $1200^{+300}_{-1200}$ & ~ & $-550^{-1200}_{+170}$\\
low-$\xi$ Emi. $\Omega$ & $1.2^{+0.7}_{-0.6}$ & ~ & $1.0^{+0.7}_{-0.7}$ \\
\hline
Milky Way ${\rm N}_{\rm H}~ (10^{22}~{\rm cm}^{-2})$ & ~ & $8.9^{+0.1}_{-0.1}$  \\
\hline
Luminosity $(10^{37}~{\rm erg}~{\rm s}^{-1})$ & $6.00\pm0.02$ & ~ & $6.17\pm0.02$ \\
\hline
\end{tabular}
\vspace*{\baselineskip}~\\
\end{center} 
\caption{Best-fit continuum and photoionization model parameters for 4U~1630$-$472.  The components are listed from top to bottom to trace the path of emitted radiation.  The continuum component parameters are listed first, followed by the absorption zones within the system, and finally absorption within the Milky Way.  
The absorption (and, coupled re-emission) zones were modeled using ``pion,'' and radially layered in the order indicated in the table (roughly in order of velocity shift).  Parameters listed between the phases were linked in joint fits.  Parameters lacking errors were fixed at the value listed.
The UFO zone describes a potential ultra-fast outflow, evident as absorption at 7.15~keV.  The high-$\xi$ zone describes the relatively constant ionized absorption that is dominated by Fe XXVI close to 6.97~keV; this zone did not require re-emission.  The mid-$\xi$ zone describes the highly variable absorption that is dominated by the Fe XXV close to 6.67 and 6.70~keV.  The low-$\xi$ zone describes weak absorption near to Fe XXV that may shift its velocity dramatically.  Over the 2.4--10.4~keV fitting band, this model achieves a Cash statistic of $C = 5424.3$ for $\nu = 5384$ degrees of freedom.  The luminosity is quoted in the ionizing band (13.6~eV to 13.6 keV) using the continuum emission components.  Please see the text for additional details.} 
\end{table}

\end{document}